\documentclass[pra, twocolumn, showpacs]{revtex4}

\usepackage{mathptmx, amsmath, amsthm, amssymb, graphicx, setspace, hyperref}
\pdfoutput=1

\newcommand{\ket}[1]{|#1\rangle}

\newtheorem*{thm}{Theorem}

\newcommand{\arxiv}[1]{\href{http://arxiv.org/abs/#1}{\tt arXiv:\nolinkurl{#1}}}

\begin{document}

\title{Faster transport with a directed quantum walk}
\date{\today}
\author{Stephan Hoyer}
\thanks{Author to whom correspondence should be addressed}
\altaffiliation[Present address: ]{Department of Physics, University of California, Berkeley, CA 94720}
\email{shoyer@berkeley.edu}
\affiliation{Department of Mathematics, University of California, San Diego, CA 92093}
\author{David A. Meyer}
\email{dmeyer@math.ucsd.edu}
\affiliation{Department of Mathematics, University of California, San Diego, CA 92093}

\begin{abstract}
We give the first example of faster transport with a quantum walk on an inherently directed graph, on the directed line with a variable number of self-loops at each vertex. These self-loops can be thought of as adding a number of small dimensions.
This is a discrete time quantum walk using the Fourier transform coin, where the walk proceeds a distance $\Theta(1)$ in constant time compared to $\Theta(1/n)$ classically, independent of the number of these small dimensions. The analysis proceeds by reducing this walk to a walk with a two dimensional coin.
\end{abstract}


\maketitle

\section{Introduction}

A quantum random walk is the quantum analog of a classical random walk where the walker may now exist in a superposition of locations.
These quantum walks offer improvement in various measures of speed over corresponding classical versions, and they can be run efficiently on quantum computers. Accordingly, they form the basis for a family of quantum algorithms that are the accelerated version of classical randomized algorithms~\cite{Ambainis2003, Kempe2003}, and provide one of few alternatives to algorithms based on Grover's search or Shor's factoring algorithm. Significant examples include structured search~\cite{Shenvi2003} and exponential speedup solving a randomized graph problem~\cite{Childs2003}.
To date, all applications of quantum walks have all been on undirected graphs, but many classical random walk algorithms of interest use directed graphs, such as classes of Markov chain Monte Carlo simulations and algorithms for fast solving of satisfiability problems~\cite{Schoning1999}. It would be very significant to accelerate these algorithms using quantum walks.

Quantum walks on arbitrary undirected graphs can be defined in both continuous and discrete time. In continuous time, a Hamiltonian $H$ is given by the connectivity matrix of the graph, so unitary evolution is given by the operator $e^{i H t}$.
In discrete time, additional degrees of freedom are generally added by associating an independent state to every distinct vertex and adjacent edge pair $\ket{v, e}$, where $e$ is an edge connecting $v$ to $v^\prime$~\cite{Watrous1999}. If the number of adjacent edges is constant across vertices, then this is equivalent to the direct product of vertex space $\mathcal{H}_v$ and a ``coin space'' $\mathcal{H}_c$. In this case there are two operations associated with the walk, a coin $C$ and shift $S$. $C$ performs some unitary operation in the coin space $\mathcal{H}_c$, and $S$ performs the shift $\ket{v, e} \to \ket{v^\prime, e}$.  The evolution of the walk is given by repeated application of the operation for a single time step, $U = S \cdot (I \otimes C)$. In the case of general undirected graphs, a generalized coin replacing $I \otimes C$ may have a different action at each vertex, but still operates independently on the subspaces corresponding to each vertex.

Directed quantum walks have been considered previously, but unitary evolution on arbitrary directed graphs requires reversibility that is not always available.
Montanaro has shown that a discrete time directed walk may be constructed for any graph where it is possible to return to every position by associating coins with each cycle in the graph, and suggests making graphs irreversible by partial measurement of regions~\cite{Montanaro2007}. This is a prescription for a quantum walk but under continuous measurement it no longer corresponds to the classical walk. This walk also requires global knowledge of the graph and thus seems unlikely to be useful for algorithms, as in most cases this is equivalent to already knowing the answer to the problem the graph represents.

In contrast, following Severini~\cite{Severini2003} we identify an arbitrary quantum walk on a directed graph by associating each edge of the graph with an independent state, very similarly to the general prescription for undirected graphs.
Then any pairing between incoming and outgoing vertices on a graph where each vertex has the same number of edges in and out will allow for unitary evolution.
This condition on suitable graphs is clearly sufficient; it can also be shown to be necessary~\cite{Severini2003}.
The coin operator $C$ applies some unitary transformation in the subspace associated with the source of each edge at a vertex. Then the shift operator $S$ transfers the amplitudes associated with each edge incoming to a vertex to some edge outgoing from the same vertex. Implicit in the choice of shift operator is some choice of pairings between incoming and outgoing edges. In undirected walks this is done by default, by pairing incoming edges with themselves, but in the more general case of directed walks this option is not always available.

Our objective is algorithmic speedup, but before we can expect speedup using a directed walk to yield an algorithm, precedence suggests that we should first identify a toy example to show that such speedup is possible.
For undirected walks, this example is the walk along the line, where the expectation value of the position increases as $\Theta(t)$ instead of $\Theta(\sqrt{t})$ classically~\cite{Ambainis2001}.
However, as yet there have been no concrete examples of quantum walks on fundamentally directed graphs that are in any sense faster than their classical equivalents.
We present the first such directed walk, with an improvement in the expectation value of position along a directed line in constant time to a constant $\Theta(1)$ from $\Theta(1/n)$ classically, where $n$ indicates the dimensionality of the graph.

\section{Directed walk with self-loops}

\begin{figure}[]
	\includegraphics[]{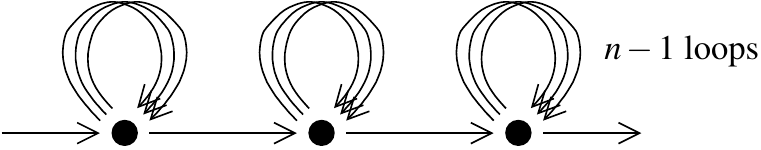}
	\caption{\label{fig:linewithloops}Directed walk on the line with $n-1$ self-loops at each vertex.}
\end{figure}

Consider a directed random walk along a line with $n-1$ self-loops at each vertex, as shown in Fig.~\ref{fig:linewithloops}.
Alternatively, we can consider these self-loops as paths around an infinitesimally small $n-1$ dimensional spatial torus $T$ and the graph as a discretization of $\mathbb{R} \times T$.
Initialized at a vertex $x=0$, classically there is a probability $1/n$ of moving to the next vertex, so the expected position $\langle x \rangle = t/n$ after $t$ steps, which is $\Theta(1/n)$.
This graph is necessarily more complicated than the base case for undirected walks, because simple classical directed walks oriented in one direction already proceed a distance $\Theta(t)$ leaving no room for improvement, unlike $\Theta(\sqrt{t})$ for the undirected walk. Accordingly we seek a different kind of speedup.

In the quantum version of this walk, we assign a basis vector to each edge on our graph, identifying each edge with the vertex at its base. Thus any state is a linear combination of the states
\begin{align}
\ket{x, \rightarrow}, \ket{x, 1}, \ket{x, 2}, \ldots, \ket{x, n-1},
\label{eq:x-basis}
\end{align}
for all non-negative integer values of $x$, where $\rightarrow$ indicates the edge along the line and each number indicates a distinct self-loop. Recall the $n$-dimensional discrete Fourier transform may be written as
\begin{align}
C = \frac{1}{\sqrt n} \left( \begin{matrix}
1 & 1 & 1 & \cdots & 1 \\
1 & \omega & \omega^2 & \cdots & \omega^{n-1} \\
1 & \omega^2 & \omega^4 & \cdots & \omega^{2(n-1)} \\
\vdots & \vdots & \vdots & \ddots & \vdots \\
1 & \omega^{n-1} & \omega^{2(n-1)} & \cdots & \omega^{(n-1)(n-1)} \\
\end{matrix} \right),
\label{eq:dftmatrix}
\end{align}
where $\omega = e^{2\pi i/n}$ is the $n$th root of unity.
Then one step of our quantum walk is given by applying $C$ as a coin operator on the coin space in the basis given by Eq.~\eqref{eq:x-basis}, and then applying the shift operator $S$ transferring amplitudes from incoming edges at a vertex to outgoing edges.
These edges are paired by the shift operator in the only non-symmetry breaking manner, by associating self-loops with themselves and pairing the incoming and outgoing paths along the line,
\begin{subequations}
\begin{align}
	S \ket{x, \rightarrow} &= \ket{x+1, \rightarrow} \\
	S \ket{x, k} &= \ket{x, k}.
\end{align}
\end{subequations}

\begin{figure}[]
	\includegraphics[]{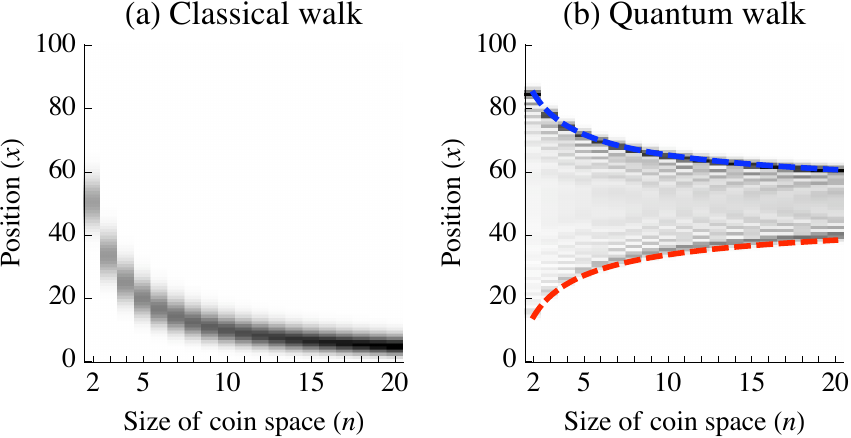}
	\caption{\label{fig:walkplot}(Color online) Probability of measurement at each position after 100 time steps as a function of the size of the coin space, for (a)~classical and (b)~quantum walks on the graph shown in Fig.~\ref{fig:linewithloops}. The boundaries of the interval for the quantum walk given by Eq.~\eqref{eq:interval} are indicated by dashed lines.}
\end{figure}

We initialize this walk in the state $\ket{0,\rightarrow}$ and repeatedly apply our time evolution operator. As with many other such proofs with quantum walks, our proof of its performance relies on reduction to the quantum walk on the line.
The performance gains from this quantum walk can also be verified numerically, as shown in Fig.~\ref{fig:walkplot}.

\begin{thm}
This quantum walk proceeds an expected distance $\Theta(1)$ in constant time independent of the number of self-loops.
\begin{proof}
Consider the action of one iteration of the time evolution operator $U = S \cdot (I \otimes C)$ on the initial state $\ket{x, \rightarrow}$,
\begin{align}
U \ket{x, \rightarrow} = {\scriptstyle \frac{1}{\sqrt n}} \ket{x + 1, \rightarrow} + {\scriptstyle \sqrt\frac{n-1}{n}} \ket{x, \circlearrowleft},
\label{eq:arrow-action}
\end{align}
where as the amplitude on each self-loop is the same, they are grouped into a single normalized state
\begin{align}
\ket{x, \circlearrowleft} = \frac{1}{\sqrt{n-1}} \sum_{k=1}^{n-1} \ket{x, k}.
\end{align}
Now consider the action of $U$ on the superposition of loops state $\ket{x, \circlearrowleft}$. It is given by
\begin{align}
U \ket{x, \circlearrowleft} = {\scriptstyle \sqrt\frac{n-1}{n}} \ket{x + 1, \rightarrow} - {\scriptstyle \frac{1}{\sqrt n}} \ket{x, \circlearrowleft},
\label{eq:loop-action}
\end{align}
as each non-zero power of $\omega$ in the Fourier transform is a root of the polynomial
\begin{align*}
x^{n-1} + \ldots + x + 1 = \prod_{k=1}^{n-1} (x - \omega^k)
\end{align*}
and thus for all $k$,
\begin{align*}
\omega^k + (\omega^k)^{2} + \ldots + (\omega^k)^{n-1} = -1.
\end{align*}

Equations \eqref{eq:arrow-action} and \eqref{eq:loop-action} reduce this walk to a quantum walk along the line with the two dimensional coin
\begin{align}
C^\prime = \left( \begin{matrix} \beta & \alpha \\ \alpha & -\beta \end{matrix} \right), \qquad \text{where } \alpha = \sqrt\frac{n-1}{n}, \quad \beta = \frac{1}{\sqrt n} ,
\end{align}
on the coin space given by $\{ \rightarrow, \circlearrowleft \}$.
Such a walk can be easily shown to be equivalent to the standard quantum walk on the line with the coin space $\{R, L\}$ under a simple map and thus this walk proceeds similarly in linear time.
But as Bressler and Pemantle have proven directly with this coin space~\cite{Bressler2007}, the form of the coin matrix $C^{\prime}$ immediately specifies the probability distribution as spread over the interval
\begin{align}
\left[ \frac{1 - \beta}{2} t, \frac{1 + \beta}{2} t \right],
\label{eq:interval}
\end{align}
where the probability of being found outside the interval decays at least as fast as $t^{-N}$ for any integer $N>0$. Figure~\ref{fig:walkplot}(b) shows the bounds of this interval. Thus this walk proceeds a constant distance $\Theta(1)$ in constant time independent of the number of loops $n-1$, and as $n\to\infty$ we have $\beta \to 0$, so $x \sim t/2$.
\end{proof}
\end{thm}

As a caveat, the speed of this walk depends completely on the natural pairing of
incoming and outgoing edges. If these edges are paired randomly at each
vertex, this quantum walk slows down to the classical speed. In algorithmic terms, this means the walk requires some local sense of where the solution is in order to
proceed faster. This makes the improvement in speed with this walk less remarkable, as such information would not be accessible for use in search algorithms, to date the most significant direct algorithmic use of quantum walks.

\section{Further directed quantum walks}

The speedup found here is certainly different from that in the classic examples for undirected quantum walks, but it may still be of note. The scaling of speed with the number of dimensions is also not unprecedented, as seen for instance by performance gains with quantum walks on the binary hypercube, a variation of which is the graph searched by random walk satisfiability algorithms.

Numerical investigations suggest that similar results hold even when the loops are given a finite non-trivial discretization, and faster transport is also easy to find on other directed graphs constructed with natural edge pairings. The primary obstacle to finding algorithmic applications of these directed quantum walks is most likely that graphs representing real problems are unlikely to have enough symmetry to allow such meaningful pairing of edges. Since quantum walks such as this one rely on symmetry for useful quantum interference, this challenge may remain significant.

Also remaining is the challenge of constructing useful quantum walks on directed graphs that remain fundamentally irreversible. In this case, at vertices where there are more paths in than out, some sort of decoherent process must be used to reduce the dimensionality of the space in which the walker moves. This may framed in terms of partial measurement of a path, or more broadly as a general quantum operation. Whether quantum speed-up could still be preserved through such operations remains unclear.

\begin{acknowledgments}
SH was partially supported by NSF Research Experience for Undergraduates Grant No.\ PHY-0552402. DM was partially supported by ARO Grant No.\ W911NF071003. \end{acknowledgments}

\bibliography{directedwalks}

\end{document}